\documentclass[a4paper]{jpconf}
\usepackage{graphicx}
\usepackage{wrapfig}

\usepackage{amsfonts,amssymb,amsmath}

\newcommand{\be}{\begin{equation}}
\newcommand{\ee}{\end{equation}}
\newcommand{\bea}{\begin{eqnarray}}
\newcommand{\eea}{\end{eqnarray}}
\newcommand{\beal}{\begin{aligned}}
\newcommand{\eeal}{\end{aligned}}

\begin{document}
\title{Accelerating Black Holes}

\author{Ruth Gregory}

\address{Centre for Particle Theory, Durham University,
South Road, Durham, DH1 3LE, UK\\
Perimeter Institute, 31 Caroline Street North, Waterloo, 
ON, N2L 2Y5, Canada}

\ead{r.a.w.gregory@durham.ac.uk}

\begin{abstract}
In this presentation, I review recent work \cite{Appels:2016uha,Appels:2017xoe}
with Mike Appels and David Kubiz\v n\'ak on thermodynamics of accelerating
black holes. I start by reviewing the geometry of accelerating black holes, 
focussing on the conical deficit responsible for the `force' causing the black hole
to accelerate. Then I discuss black hole thermodynamics with conical deficits,
showing how to include the tension of the deficit as a thermodynamic variable,
and introducing a canonically conjugate {\it thermodynamic length}. Finally, I
describe the thermodynamics of the slowly accelerating black hole in anti-de
Sitter spacetime.
\end{abstract}

\section{Introduction}

Black holes have proved to be perennially fascinating objects to study, both
from the theoretical and practical points of view. We have families of exact
solutions in General Relativity and beyond, and an ever better
understanding of their phenomenology via numerical and observational
investigations. However, in all of these situations the description of the
black hole is as an isolated object, barely influenced by its sometimes
extreme environment, its only possible response being to grow by accretion.
The familiar exact solutions in GR describe precisely this -- an isolated,
topologically spherical, solution to the Einstein equations.
The black hole is the ultimate slippery object -- to accelerate,
we must be able to `push' or `pull' on the actual event horizon, yet, by
its very nature, anything touching the event horizon must be drawn in, unless it
travels locally at the speed of light. Translating this to a physical object 
touching the event horizon, this means that the energy momentum tensor 
must locally have the form $T^0_0 \sim T^r_r$. Luckily, we have a candidate
object that satisfies this requirement.

A cosmic string \cite{Vilenkin:1984ib} is a very thin, quasi-linear object, with 
an energy momentum
dominated by its mass per unit length, and a string tension equal in magnitude 
to this energy. Cosmic strings can be formed in field theories with non-simply
connected vacua, and gravitationally produce no long range local curvature,
but generate an overall global conical deficit in the spatial sections orthogonal 
to the string, thus a string, effectively, is a local conical deficit in the spacetime.
It turns out that this conical deficit is precisely what accelerates the black hole.

\section{Accelerating black holes}

The accelerating black hole metric has been known for many years, though
perhaps less widely outside the classical GR community, thus it is worth briefly
reviewing its structure. It is described by the {\it C-metric} 
\cite{Kinnersley:1970zw,Plebanski:1976gy,Griffiths:2005qp}, 
which can be written in Schwarzschild-like coordinates \cite{Hong:2003gx} as
\be
ds^2=\frac{1}{(1+A r \cos\theta)^2}\Biggl\{ 
f(r) dt^2
-\frac{dr^2}{f(r)} - r^2 \left[ \frac{d\theta^2}{g(\theta)} 
+ g(\theta)\sin^2\theta \frac{d\phi^2}{K^2}\right]\Biggr\}
\label{cmet}
\ee
where $\phi$ has periodicity $2\pi$, and 
\be
f(r)=(1-A^2r^2)\Bigl(1-\frac{2 m}{r}\Bigr)
+\frac{r^2}{\ell^2} \;,\qquad
g(\theta)=1+2mA \cos\theta\,,
\label{fandg}
\ee
where $m$ represents the mass scale of the black hole, $A$ the
acceleration, $K$ the conical deficit,
and $\ell$ the (negative) cosmological constant, 
via $\Lambda = -3/\ell^2$. It is worth briefly deconstructing the
C-metric to highlight the importance of the various parameters
to the geometry and physics of the solution.
\begin{wrapfigure}{r}{0.45\textwidth}
\begin{center}
\includegraphics[width=0.43\textwidth]{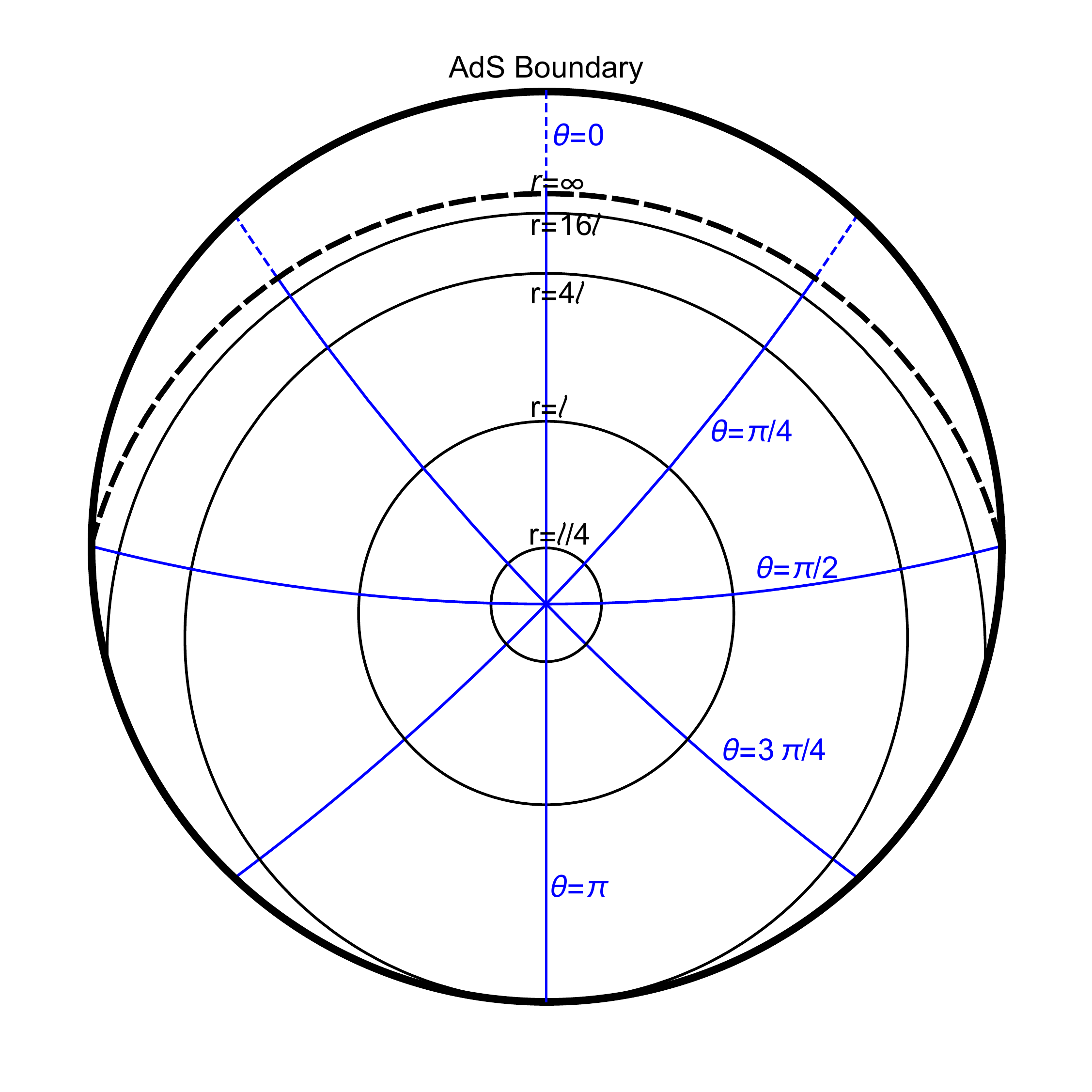}
\end{center}
\caption{\label{fig:slowacc}AdS spacetime in slowly accelerating coordinates.}
\end{wrapfigure}

The function $f(r)$ determines the horizon structure of the black 
hole; whether there are one, two, or no horizons, depends 
on the mass parameter and the competition between the acceleration
and cosmological constant. First, to understand the role of $K$, set
$A\to0$. Then \eqref{cmet} is clearly identifiable as the Schwarzschild-AdS
solution, with the proviso that $\phi$ is now divided by $K$. Focussing
on the axis through the black hole shows that that circles around this 
axis have a circumference $2\pi (1-1/K)$ times their proper radius,
instead of the expected $2\pi$ of a locally flat space. Thus, the axis 
contains a conical deficit 
\be
\delta = 2\pi \left ( 1 - \frac1K \right) = 8\pi \mu
\label{mudef}
\ee
running through the black hole \cite{Aryal:1986sz} that can be replaced 
by a non-singular cosmic string \cite{Achucarro:1995nu}.

Now consider the acceleration parameter, $A$. This in part counterbalances
the effect of the cosmological constant in $f(r)$, and if $m$ is nonzero, gives
a distortion of the two-spheres at constant $r$. To isolate the effect of $A$,
setting $m=0$ and $K=1$ in \eqref{cmet} and keeping $A\ell<1$ means
that there is no acceleration horizon, and a simple coordinate transformation
\be
1 + \frac{R^2}{\ell^2} = \frac{1 + (1-A^2\ell^2)r^2/\ell^2}{(1-A^2\ell^2)\Omega^2}
\qquad;\qquad
R \sin\Theta = \frac{r\sin\theta}{\Omega}
\ee 
gives pure AdS spacetime \cite{Podolsky:2002nk}. 
The C-metric coordinates in \eqref{cmet} 
represent an off-centre perspective of AdS -- they are the coordinates of
an observer hovering at fixed distance from the centre of AdS, who is
actually accelerating (see figure \ref{fig:slowacc}), but not sufficiently
strongly to form an acceleration horizon. 

Putting together, $A$ gives the acceleration of the black hole that 
counterbalances the effect of the cosmological constant in the Schwarzschild 
potential. The acceleration also gives an imbalance between the North 
and South axes that now have different conical deficits:
\be
\beal
&\bullet \;\; \theta\to\theta_+=0 \quad
ds_{\theta,\phi}^2 \propto d\theta^2 + {(1+2mA)^2 \over K^2} \theta^2 d\phi^2 \\
&
\bullet \;\; \theta\to\theta_-=\pi \quad
ds_{\theta,\phi}^2 \propto d\theta^2 + {(1-2mA)^2 \over K^2} (\pi -\theta)^2 d\phi^2
\eeal
\ee
leading to the interpretation of the acceleration being induced by 
the imbalance of tension between the North and South axes:
\be
8\pi \mu_\pm = \delta_\pm = 2\pi \left ( 1 - \frac{g(\theta_\pm)}{K} \right)
= 1 - \frac1K \left (1 \pm 2mA + e^2A^2 \right)
\label{tensions}
\ee
\begin{wrapfigure}{r}{0.4\textwidth}
\begin{center}
\vskip -1cm
\includegraphics[width=0.25\textwidth]{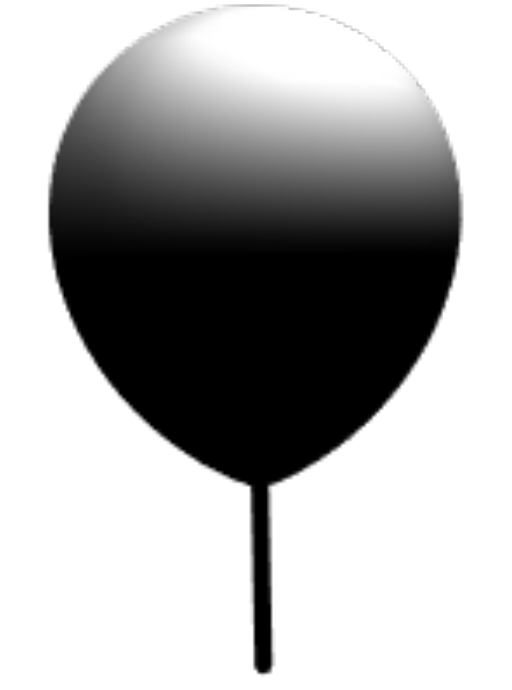}
\end{center}
\caption{\label{fig:accbh}A representation of the accelerating black hole.}
\end{wrapfigure}
Conventionally, we choose to make the N axis regular by setting $K=1+2mA$, 
leaving the deficit on the S axis corresponding to a cosmic string of tension 
$\mu_S = mA/K$ \cite{Gregory:1995hd} (see figure \ref{fig:accbh}. 
We will also demand that
we have a {\it slowly accelerating} black hole, i.e.\ the acceleration 
is sufficiently gentle that we do not get an acceleration horizon (almost 
equivalent to $A\ell<1$) \cite{Krtous:2005ej}. The geometry therefore
looks like a black hole in AdS displaced from the centre, the
force required to hold it being supplied by the cosmic string or 
conical deficit running from the horizon to the boundary.
Interpreting this tension as the force, and $A$ as the acceleration, 
Newton's Law suggests that the mass\footnote{Note that
unlike the rapidly accelerating black hole, this is a genuine
ADM-style mass, and not a ``re-arrangement of dipoles''
as discussed in \cite{Dutta:2005iy} where a boost mass
was introduced.} of the black hole should 
then be identified as $M=m/K$, so that $\mu = F =  MA$ --
we will see presently that this is exactly what the first law of
thermodynamics requires.

\section{Thermodynamics with conical defects}

Before discussing the accelerating black hole, set $A=0$ and
look at the effect on the Schwarzschild-AdS black hole of the conical 
deficit (see \cite{Martinez:1990sd} for early work with fixed tension deificits). 
The horizon is defined as $f(r)=0$, 
so we examine incremental changes in the black hole
\be
f(r_+ + \delta r_+) = f'(r_+) \delta r_+ + \frac{\partial f}{\partial m} 
\delta m + \frac{\partial f}{\partial \ell} \delta \ell = 0
\label{varyf}
\ee
where we are allowing not only for a change in the black hole 
mass, but a possible change in the cosmological constant, as might
occur for example when vacuum energy is due to a scalar field
that is allowed to vary (see \cite{Gregory:2017sor} for a discussion
of black hole thermodynamics in this situation).

Defining temperature in the usual way via surface gravity,
$T = f'(r_+)/4\pi$, we see that the first term in \eqref{varyf} 
will be related to a change in entropy, however, now the entropy,
given by the area of the horizon \cite{Herdeiro:2009vd}, also depends on $K$:
\be
S = \frac{\pi r_+^2}{K} \rightarrow
f'(r_+)\delta r_+ = \frac{2K}{r_+} \left ( T\delta S 
+ \frac{ r_+^2 f'(r_+)}{4} \frac{\delta K}{K^2} \right ) 
\ee
To identify the physics in changing $K$, note that tension was related
to $K$, \eqref{mudef}, from which we deduce
\be
\delta \mu = \frac{\delta K}{4 K^2}
\ee
Finally, writing the thermodynamic pressure 
\cite{Kastor:2009wy,Dolan:2010ha,Dolan:2011xt,Kubiznak:2012wp}
\be
P = -\Lambda = \frac{3}{8\pi\ell^2} \qquad V = \frac{4\pi r_+^3}{3K}
\ee
allows us to identify a first law from \eqref{varyf}
\be
\delta (\frac{m }{K}) = T \delta S + 2(m-r_+)\delta \mu +  V \delta P 
\ee
Leading to the definition of mass for the black hole $M = m/K$,
as suggested by Newton's Law. Indeed, taking this definition of $M$
also gives a Smarr formula \cite{Smarr:1972kt} for the black hole:
\be
M = 2TS - 2PV
\ee

Let us look more closely at this first law. The term multiplying the
variation in tension has the form of a length, and we define 
this to be the {\it thermodynamic length}\footnote{Note, a 
similar expression was obtained by Kastor and Traschen 
\cite{Kastor:2012dt} in the context of an angular gravitational
tension associated to the rotational symmetry of the black hole.} 
associated to the tension
charge of the string:
\be
\lambda = r_+ - m
\label{tdlength}
\ee
Note, we do not include the factor of $2$, but instead have the length of the 
string from each pole -- this will be particularly relevant when we look at the
accelerating black hole that has different deficits at each pole. This reinforces 
the interpretation of $M$ as the enthalpy \cite{Kastor:2009wy} of the black hole:
if the black hole grows, it swallows some string, but has also displaced the 
same amount of energy from the environment.
\begin{figure}[h]
\begin{center}
\includegraphics[width=0.8\textwidth]{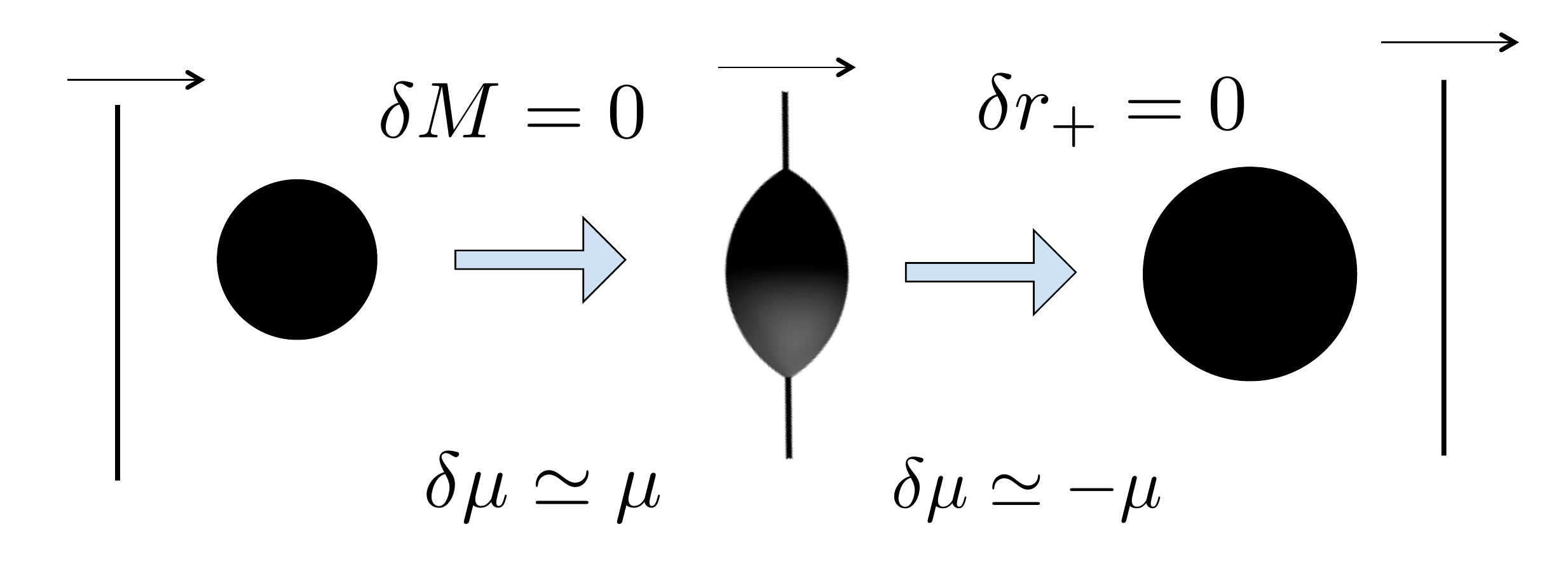}
\end{center}
\caption{\label{fig:capture}A black hole capturing a piece of cosmic string.}
\end{figure}

Does this definition and the first law make sense? In order to explore this,
consider the situation of a cosmic string interacting with a black hole,
first proposed in \cite{Bonjour:1998rf},
and for simplicity let $\Lambda=0$, for which the first law reduces to
\be
\delta M = T\delta S - 2 \lambda \delta \mu
\ee
Initially, the string is outside the black hole, but moving towards it (see figure
\ref{fig:capture}). The 
black hole then captures the string and thus acquires a positive conical
deficit, however because the string is still moving, it eventually pulls off the
black hole, leaving a purely spherical black hole behind. The first step of
the process has a positive $\delta \mu$ and the second an equal but
negative tension change, $-\delta\mu$. In the first stage, we would expect
the total energy of the black hole to be constant, thus $\delta M=0$, 
the physical intuition being that a cosmic string passing through a spherical 
shell of matter would not change the total energy throughout the process,
For the black hole however, string cannot emerge from within the event
horizon, and we have the interpretation of a segment of 
string being captured by the black hole, with the black hole increasing
its mass accordingly. 

In the first stage (taking $\mu \ll 1$), fixing $M$ implies 
\be
T \delta S =  2\lambda \mu
\ee
consistent with $\delta m = 4 m \delta \mu$, and $S = 4\pi m^2/K$.
Because the energy of the black hole has been fixed,
the event horizon moves outwards to compensate for the conical 
deficit, resulting in an increase of entropy indicating this is an irreversible
thermodynamic process. 
In the second stage, the string pulls off the black hole, so $\delta \mu = -\mu$,
and since entropy cannot decrease, $M$ must increase:
\be
\delta M = T\delta S + 2m\mu
= \delta m + 4m\mu 
\ee
in this case, it seems reasonable to suppose that the local geodesic 
congruence defining the event horizon cannot contract, giving
$\delta m=0$, and $\delta M = 4m\mu$. Thus, the net result of this
adiabatic process is that the black hole has `swallowed' the segment of
cosmic string that fell inside its event horizon.

\section{Thermodynamics of accelerating black holes}

To find the first law for accelerating black holes, the idea is to follow the
same steps: the horizon is defined by $f(r)=0$, and consider small
increments in horizon radius:
\be
\delta f(r_+) = f_+' \delta r_+  - 2 \frac{\delta m}{r_+} (1-A^2r_+^2)
- 2 A \delta A r_+ (r_+ - 2m) - 2 \frac{r_+^2}{\ell^3} \delta \ell = 0
\ee
Using the usual Euclidean method we can relate temperature to $f_+'$,
\be
T = \frac{f'(r_+)}{4\pi} = \frac{1}{2\pi r_+^2} \left [ m(1-A^2 r_+^2) 
+ \frac{r_+^3}{\ell^2(1-A^2 r_+^2)} \right ]
\label{Tacc}
\ee
but now the area of the event horizon gives a more involved relation
for $\delta r_+$:
\be
\delta S = \delta \left ( \frac{\pi r_+^2}{K(1-A^2 r_+^2)}\right)
=\frac{2\pi r_+ \delta r_+}{K(1-A^2 r_+^2)^2} + 
\frac{2 \pi r_+^4 A\delta A}{K (1-A^2 r_+^2)^2} - 
\frac{\pi r_+^2}{(1-A^2 r_+^2)} \frac{\delta K}{K^2}
\ee
The expressions for the tensions \eqref{tensions} allow
us to replace $\delta K/K^2 = 2 ( \delta \mu_+ + \delta \mu_- )$
and $m\delta A /K= -\delta \mu_+ + \delta \mu_- - A \delta M$,
having identified $M=m/K$ as before. Putting together, this gives
the First Law as (see also \cite{Astorino:2016ybm})
\be
\delta M = V\delta P + T \delta S -
\lambda_+ \delta \mu_+ - \lambda_-\delta \mu_- \,.
\label{firstaccm}
\ee
where we now have a thermodynamic length associated to
the piece of string attached at each pole:
\be
\lambda_\pm = \frac{r_+}{1 \pm Ar_+} - KM
\ee
This obviously agrees with \eqref{tdlength} for the string
threading the black hole, where $r_+$ has now been replaced by
$r_+/\Omega(r_+,\theta_\pm)$ at each pole. 
We also have a Smarr relation,
\be
M = 2 TS  - 2 PV\,,
\ee
as before. The string does not contribute to the overall mass of the black
hole as the string is gravitationally `neutral' -- it has no overall Riemann curvature,
the spacetime is locally asymptotically flat -- the effect of the string being simply
a global deficit angle.

\section{Exploring thermodynamics}

Having derived a sensible first law for the accelerating black hole,
now let us briefly explore some consequences of the expressions
(for a full discussion see \cite{Appels:2016uha,Appels:2017xoe}). 
\begin{wrapfigure}{r}{0.5\textwidth}
\begin{center}
\includegraphics[width=0.48\textwidth]{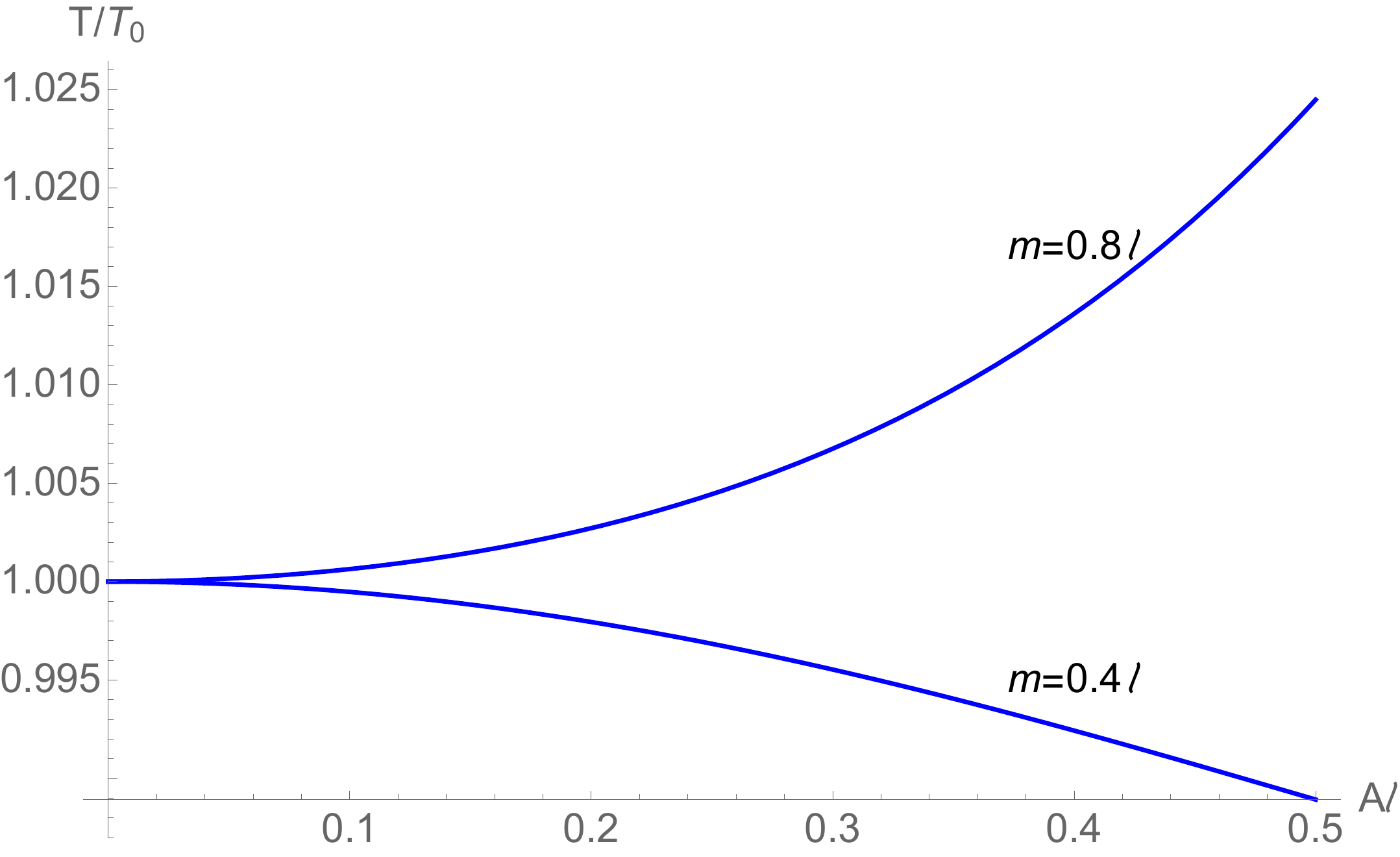}
\end{center}
\caption{\label{fig:tva}The variation of temperature of 
a fixed mass black hole with acceleration.}
\end{wrapfigure}
First consider the temperature 
of the accelerating black hole \eqref{Tacc}. Typically, we associate a
temperature to the surface gravity of an horizon; apart from the 
black hole, horizons arise in the context of cosmology with a positive 
cosmological constant, and acceleration. Here we have explicitly
taken our acceleration sufficiently low that we never get an acceleration
horizon, but nonetheless, the effect of acceleration counterbalances the
cosmological constant in the gravitational potential: $1/\ell^2 \to 1/\ell^2
- A^2 = 1/\ell_{\rm eff}^2$. We could perhaps expect that the temperature
of the black hole is somehow a combination of the inherent Hawking
temperature and the Rindler temperature, but in this case we would
expect a systematic shift with $A$, and this is not what happens. Instead,
for smaller mass black holes the temperature decreases with acceleration,
but for larger mass black holes it increases (see figure \ref{fig:tva}).
To some extent this can be understood as a result of the varying
tensions that are required to accelerate a fixed mass black hole
differently.

Another phenomenon we can explore is that of the Hawking-Page phase transition 
\cite{Hawking:1982dh}. For small black holes in AdS, the vacuum curvature
is subdominant to the local tidal curvature of the black hole, thus thermodynamics
is similar to vacuum with black holes being thermodynamically unstable. 
For black holes larger than the AdS radius however, the local tidal curvature
becomes subdominant to the vacuum curvature, and the black hole has positive 
specific heat. Thus in AdS, there is a minimum temperature for the black hole,
and below this temperature only a radiation bath can be a solution to the Einstein 
equations at finite $T$. At higher temperatures, one can have either a small or 
large black hole (or radiation bath). The Hawking-Page transition occurs at 
$T>1/\pi\ell$, where the large black hole is thermodynamically preferred to the
radiation bath, as can be seen by plotting the Gibbs free energy as a function 
of temperature.
\begin{figure}[h]
\center{\includegraphics[scale=0.58]{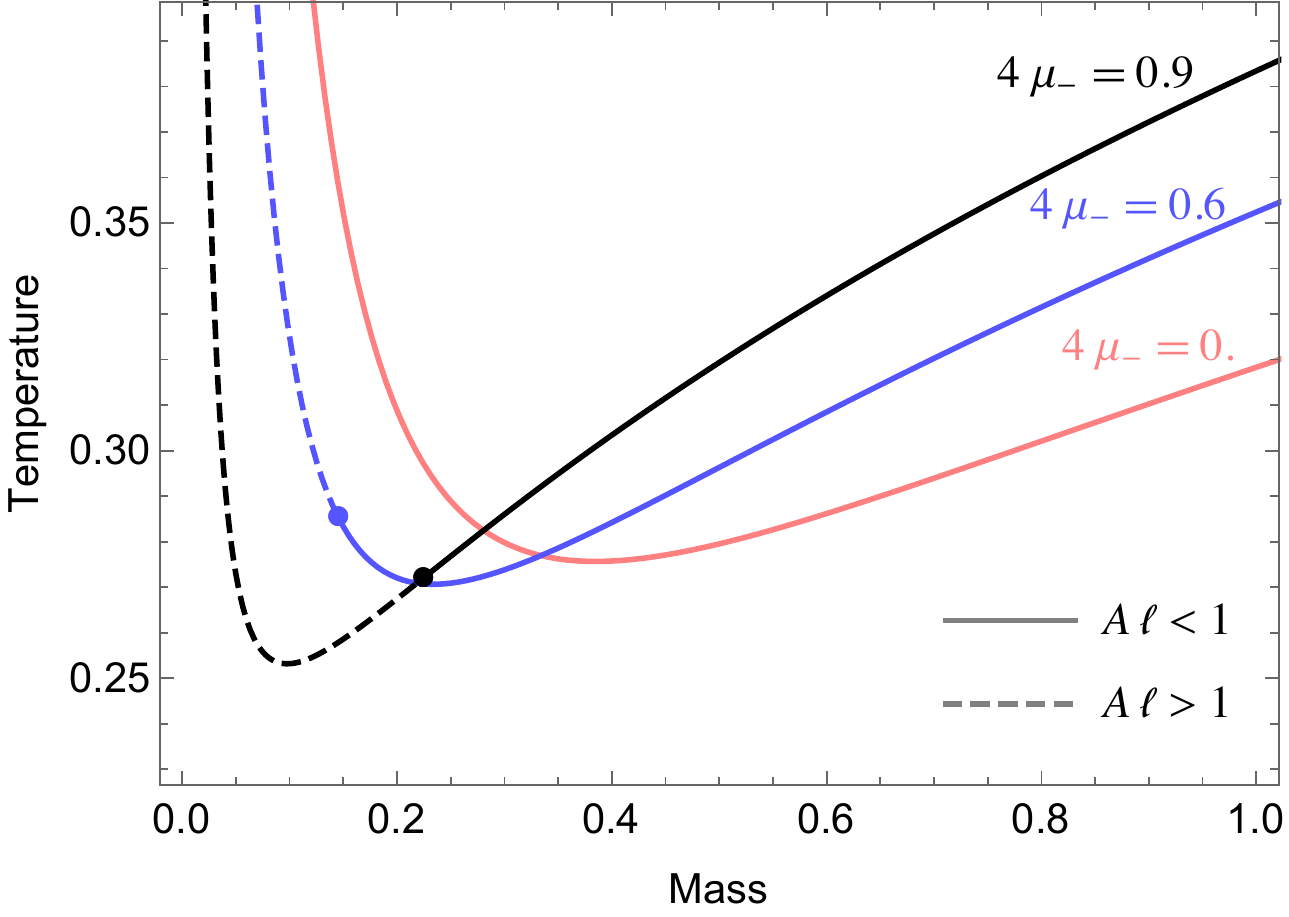}~
\includegraphics[scale=0.6]{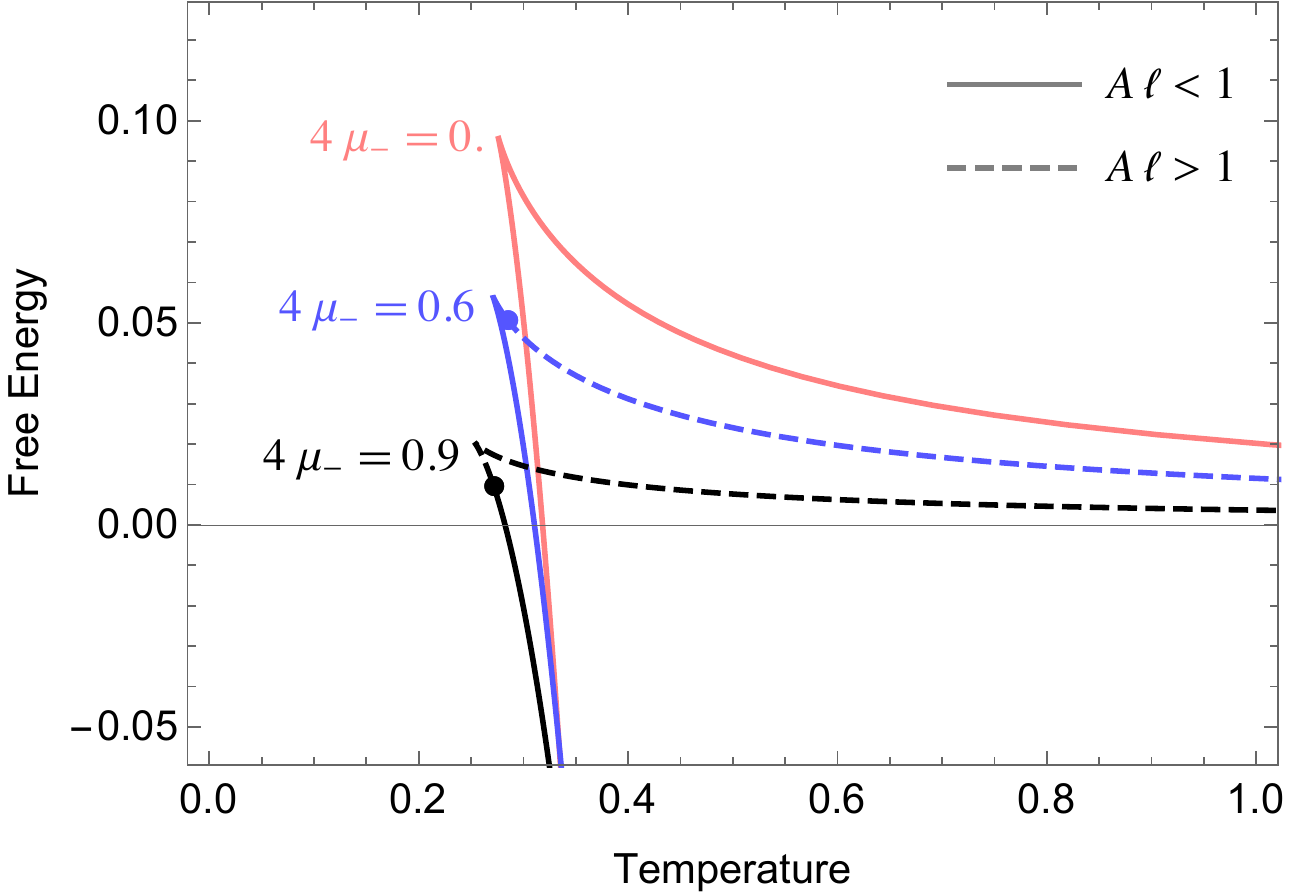}}
\caption{A plot of temperature as a function of mass (left), and
the Gibbs free energy as a function of temperature (right).
}
\label{fig:TvMnocharge}
\end{figure}
For the accelerating black hole, we fix the tension of the string, and 
plot the temperature as a function of (thermodynamic) mass  in figure 
\ref{fig:TvMnocharge}. This plot explicitly demonstrates that acceleration 
makes a black hole of given mass {\it more} thermodynamically stable 
in the sense of positive specific heat. A plot of the corresponding Gibbs free 
energy is also shown.

For simplicity, I focussed primarily in this presentation on uncharged non-rotating 
black holes to demonstrate the procedure for determining the thermodynamic
charges and potentials. One can add charge to the black hole, which now 
allows for a much richer thermodynamic phase structure. Briefly, the charges 
and potentials in the presence of electric charge become \cite{Appels:2017xoe}):
\be
\beal
M &= \frac{m}{K(1+e^2 A^2)}& Q &= \frac{e}{K}\\
\Phi &= \frac{e}{r_+} - \frac{m e A^2}{1+e^2 A^2}&
\lambda _\pm &= \frac{r_+}{1 \pm A r_+} - 
\frac{m(1-e^2 A^2)}{(1+a^2A^2)^2} \mp \frac{e^2A}{1+e^2A^2}
\eeal
\ee
with the entropy, thermodynamic pressure and volume unchanged in form.

\begin{figure}[h]
\center{\includegraphics[scale=0.58]{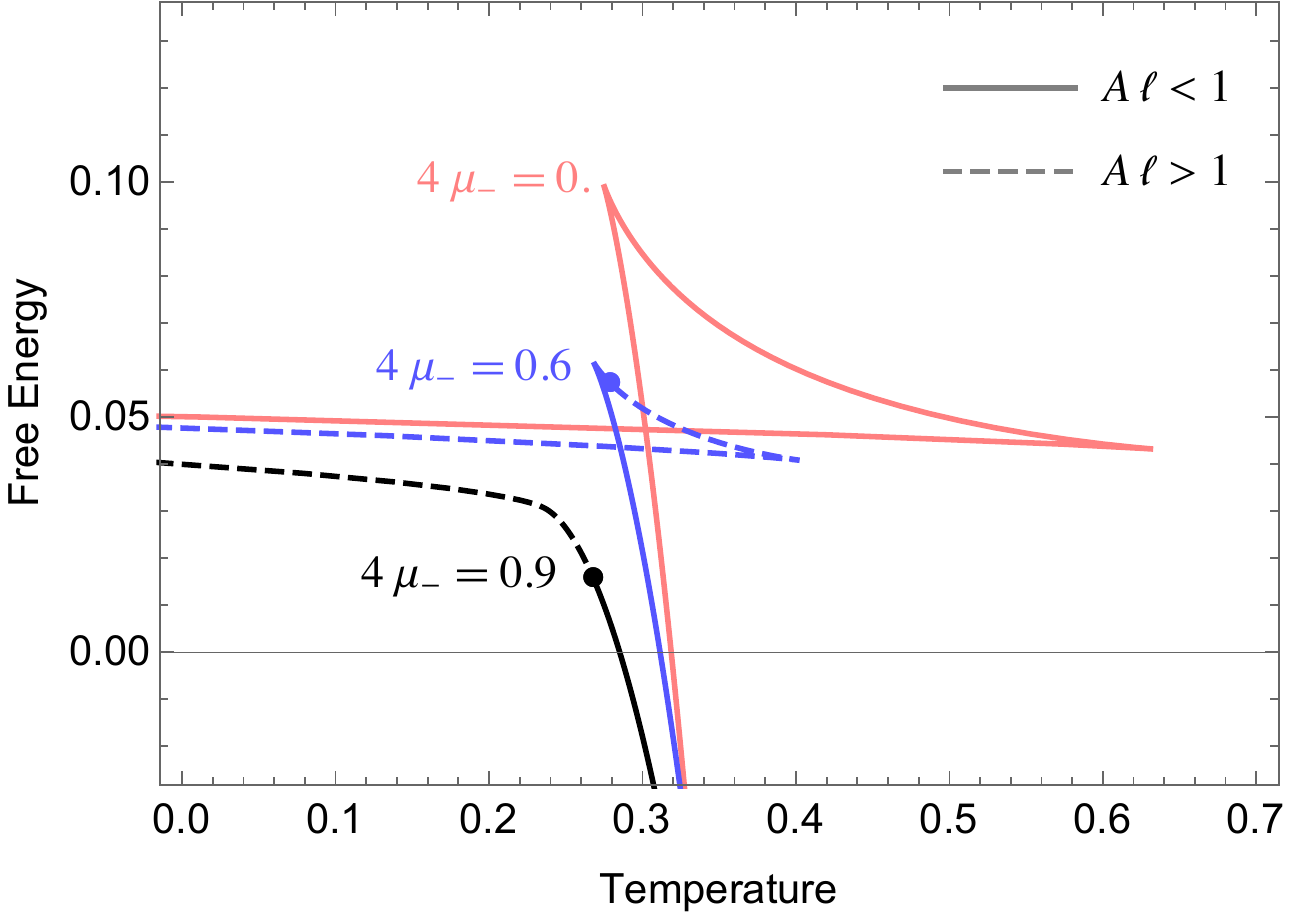}~
\includegraphics[scale=0.58]{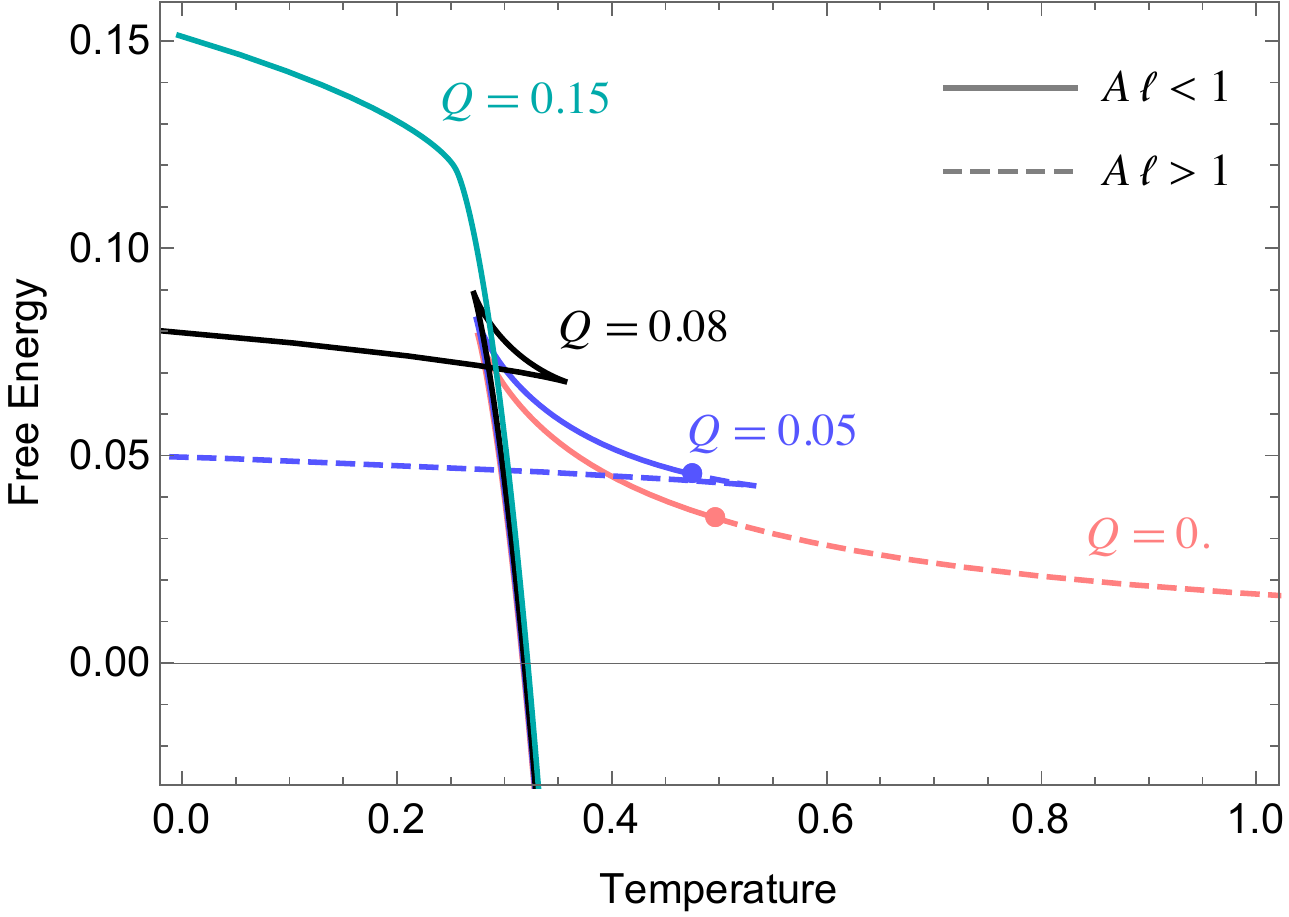}}
\caption{A plot of the free energy as a function of temperature for 
varying tension with $Q=0.05\ell$ on the left, and varying charge with $4\mu_-=0.3$
on the right. 
}
\label{fig:GvTcharge}
\end{figure}

With charge, there is a lower bound on the black hole mass from 
the extremal limit, thus at fixed $Q$, black holes have positive
specific heat near this extremal limit, and at large mass, but in between
can have negative specific heat if the charge is sufficiently low that the
event horizon lies below the AdS scale. As with the uncharged black hole,
acceleration makes the black hole more thermodynamically stable, as
seen in figure \ref{fig:GvTcharge}.

\section{Conclusion}

To sum up: we have shown how to allow for varying tension in the thermodynamics
of black holes. In accelerating black holes, if one does not allow the tension of
the deficits to vary, then the various thermodynamic charges are not free
to vary independently, and there is no guarantee that any thus derived 
thermodynamic quantities will be the correct ones, it is therefore mandatory
to consider $\mu$ as a variable in order to correctly explore thermodynamics.
Work on the thermodynamics of rotating accelerating black holes is reported
on in the poster session \cite{AGK2}.

\ack

I would like to thank David Kubiz\v n\'ak and Mike Appels for collaboration on this
project. This work was supported by STFC (Consolidated Grant ST/L000407/1),
and also by the Perimeter Institute for Theoretical Physics. 
Research at Perimeter Institute is supported by the Government of
Canada through the Department of Innovation, Science and Economic 
Development Canada and by the Province of Ontario through the
Ministry of Research, Innovation and Science.

\end{document}